\documentclass[prc,twocolumn,showpacs,amsmath,amssymb]{revtex4}
\usepackage{rotating}
\usepackage{times}
\usepackage[pdftex]{hyperref}
\usepackage{layout}
\usepackage{textcomp}
\usepackage{lscape}
\usepackage{graphicx}
\usepackage{dcolumn}
\usepackage{bm}
\usepackage{amstext}
\usepackage{amssymb,amsmath}
\usepackage{graphicx}

\begin{document}

\title{Gravitational wave from rotating neutron star}
 
\author{Shailesh K. Singh$^{1}$,  S. K. Biswal$^{1}$, M. Bhuyan$^{1}$, 
T. K. Jha$^{2}$ and S. K. Patra$^{1}$}

\affiliation{
$^{1}$Institute of Physics, Bhubaneswar-05, India. \\
$^{2}$Department of Physics, BITS Pilani, K K Birla Goa.
}
\date{\today}

\begin{abstract}
Using the nuclear equation of states for a large variety of relativistic and
non-relativistic force parameters, we calculate the static and rotating masses 
and radii of neutron stars. From these equation of states, we also evaluate the
properties of rotating neutron stars, such as rotational and gravitational 
frequencies, moment of inertia, quadrupole deformation parameter, rotational 
ellipcity and gravitational wave strain amplitude. The estimated gravitational 
wave strain amplitude of the star is found to be $\sim 10^{-23}$.

\end{abstract}
\maketitle

\section{Introduction}

Gravitational wave is a fundamental feature of elliptically deformed 
pulsars. It is produced due to the axially asymmetry of the system. 
This is one of the unique source of informations 
which can resolve most of the mystery of the stellar objects. 
About 96 percent
mass-energy of the Universe has no charge, so the major fact about
Universe can be revealed from the graviational wave (GW). However, it
is unlike to detect easily as it is done in electromagnetic wave, 
which is originated
from charge sources \cite{satya09}. Thus, the detection of gravitational
waves is  hard  due to  its low frequency and background sources.

There are experimental set up using
ground based detectors, which are specially designed for 
the measurement of gravitational waves amplitude, such as the Laser 
Interferometer 
Gravitational wave Observatory science collaboration (LIGO) and the 
German-British
Gravitational wave Detectors (GEO600). The sensitivity of the detection
is increased by the merger of these data and the upper limit of the
gravitational wave amplitude is observed to be $\sim$ $2.6\times10^{-25}$
for pulsar PSR I1603-7202 and the ellipticity of pulsar PSRI2124-3358 is 
found to be less than $10^{-6}$ \cite{prd07}.  
Some other spaced based detectors 
like Laser Interferometer Antenna (LISA) is designed
for detecting the the low frequency (0.03 mHz to 0.1 Hz) gravitational 
waves \cite{lisa} and space based Cosmic Visions 2015-2025 is in plan to 
orbiting the sun like LISA to gain more sensitiveness towards the low frequency 
gravitational wave signals \cite{satya09}. 

The neutron star (NS) and black holes are formed from the gravitational 
collapse of a highly evolved start or core collapse of an accreting white dwarf.
The neutron star is the final stage of the evolving star and then it fails to 
collapse and form a black hole due to gravity. Rotating deform neutron star
emits gravitational waves which carry the information about the 
neutron star (NS). Therefore, it is very important to discuss the upper limit 
of GW amplitude, rotational frequency $\nu_r$, quadrupole moment $\Phi_{22}$ 
and ellipticity ${\epsilon}$ of a neutron star predicted by various 
theoretical models. 
The static mass of the NS compared with recently observed data 
\cite{demorest10},
which is quite massive than the earlier measured mass from the neutron star 
pulsar PSR 1913+16 ($M=1.144M_\odot$) \cite{taylor89}.
Those equation of states (EOS) give the mass of Taylor et al. 
\cite{taylor89} fails
to reproduce the maximum mass of $(1.97 \pm 0.04)M_\odot$ \cite{demorest10}.
Thus, to get a larger mass, one needs a stiff EOS, which again oppose the 
softer EOS of kaon production \cite{kaon,transport}. To make such a model in the same
footing, extra interactions are needed as it is done in the construction
of G1 and G2 parametrizations \cite{fu96,g,mu96}. In the present paper, we have
used 20 different force parameters for both non-relativistic and relativistic mean 
field equation of states (EOS) to calculate the gravitational wave strain amplitude
of rotating neutron stars.

The paper starts with a short introduction in Sec. I. The formalisms of 
Skyrme Hartree-Fock (SHF) and Relativistic Mean Field (RMF) theory are 
presented in Sec.II. In this section we have outlined the
Hamiltonian, Lagrangian and equation of states (EOS) for non-relativistic and
relativistic formalisms. The SHF and RMF parameter sets are also tabulated in this 
section.  The calculated results of pressure and energy obtained from 
these forces are discussed in Sec. III. 
Here, the masses of the neutron stars and their respective radii
both in static and rotating frames are estimated and then used these observables
to estimate the gravitational wave strain amplitude. The related quantities like
rotational frequency $\nu_r$, quadrupole moment $\Phi_{22}$ and
ellipticity ${\cal \epsilon}$ of rotating neutron star also calculated.
The paper is summarized in Sec. IV.

\section{Theoretical Formalisms}

\subsection{Skyrme Hartree-Fock (SHF) method}

There are many known parametrizations of Skyrme interaction which reproduce the
experimental data for ground state properties of finite nuclei 
\cite{bender03,stone03} as well as the properties of infinite nuclear 
matter upto high density \cite{dutra12}.
The general form of the Skyrme effective interaction
can be expressed as a density functional $\cal H$ with some 
empirical parameters \cite{sly23ab,bender03,stone07}:
\begin{equation}
{\mathcal H}={\mathcal K}+{\mathcal H}_0+{\mathcal H}_3+ 
{\mathcal H}_{eff}+\cdots,
\label{eq:1}
\end{equation}
where ${\cal K}$ is the kinetic energy, ${\cal H}_0$ the zero range, 
${\cal H}_3$ the
density dependent and ${\cal H}_{eff}$ the effective-mass dependent terms, 
which are
relevant for calculating the properties of nuclear matter.
More details can be found in Refs. \cite{sly23ab,stone07,bender03}.
These are functions of 9 parameters
$t_i$, $x_i$ ($i=0,1,2,3$) and $\eta$ are given as
\begin{eqnarray}
{\mathcal H}_0&=&\frac{1}{4}t_0\left[(2+x_0)\rho^2 - (2x_0+1)(\rho_p^2+\rho_n^2)\right], \nonumber
\label{eq:2}
\\
{\mathcal H}_3&=&\frac{1}{24}t_3\rho^\eta \left[(2+x_3)\rho^2 - (2x_3+1)(\rho_p^2+\rho_n^2)\right], \nonumber
\label{eq:3}
\\
{\mathcal H}_{eff}&=&\frac{1}{8}\left[t_1(2+x_1)+t_2(2+x_2)\right]\tau \rho \nonumber \\
&&+\frac{1}{8}\left[t_2(2x_2+1)-t_1(2x_1+1)\right](\tau_p \rho_p+\tau_n \rho_n).   \nonumber
\label{eq:4}
\end{eqnarray}
The kinetic energy ${\cal K}=\frac{\hbar^2}{2m}\tau$, a form used in the 
Fermi gas model for non-interacting Fermions.  The total nucleon number 
density $\rho=\rho_n+\rho_p$, the kinetic energy density
$\tau=\tau_n+\tau_p$.

The standard form of the Skyrme effective interaction can
be expressed as \cite{brink72,bender75,sly23ab}:
\begin{eqnarray}
V_{eff}(r_1, r_2)&=&t_0(1+x_0P_{\sigma})\delta(r) \nonumber \\
&& +\frac{1}{2}t_1(1+x_1P_{\sigma})\left[{\bf P^{'2}}\delta(r)
+\delta(r){\bf P^2}\right] \nonumber\\
&&+t_2(1+x_2P_{\sigma}){\bf P^{'}}\cdot \delta(r) {\bf P} \nonumber \\
&& +\frac{t_3}{6}(1+x_3 P_{\sigma})
\left(\rho({\bf R})\right)^{\gamma}\delta(r).  \nonumber 
\label{eq1}
\end{eqnarray}
Here, $r=r_1-r_2$, $R=\frac{1}{2}(r_1+r_2)$,
$P=\frac{1}{2i}(\nabla_1-\nabla_2)$ and $\sigma=\sigma_1+\sigma_2$.
The main advantage of the Skyrme density functional is that it allows
the analytical expression for all variables explaining the infinite
nuclear matter characteristics. The general expression for the energy
per particle of asymmetric nuclear matter (ANM) in terms of energy
density $\varepsilon$ and number density $\rho$ is given by
\cite{sly23ab,dutra08}:
\begin{eqnarray}
\frac{E}{A} (Y_p, \rho)&=&\frac{\varepsilon (\rho)}{\rho}
=\frac{3}{10}\frac{\hbar^2}{2m}
\left(\frac{3\pi^2}{2}\right)^{2/3} \rho^{2/3}F_{5/3} \nonumber \\
&+&\frac{1}{8}t_0\rho\left[2(x_0+2)-(2x_0+1)F_2\right] \nonumber\\
&+&\frac{1}{48}t_3\rho^{(\sigma+1)}\left[2(x_3+2)-(2x_3+1)F_2\right]\nonumber\\
&+&\frac{3}{40}(\frac{3\pi^2}{2})^{2/3} \rho^{5/3}
\left[t_1(x_1+2)+t_2(x_2+2)\right]F_{5/3} \nonumber \\
&+&\frac{3}{80}(\frac{3\pi^2}{2})^{2/3} \rho^{5/3}
\left[t_2(2x_2+1)-t_1(2x_1+1)\right]F_{8/3}, \nonumber 
\label{eq1}
\end{eqnarray}
with the asymmetric factor,
\begin{eqnarray}
F_m (Y_p)=2^{m-1}\left[Y_p^m+(1-Y_p^m)\right]. \nonumber
\end{eqnarray}
The term $Y_p$ is adopted in place of isospin to define the
asymmetry of the infinite nuclear matter, which is common
notation in astrophysics. 
The analytical form of the pressure density can be expressed as: 
\begin{eqnarray}
P (Y_p, \rho)&=& \frac{1}{5}\frac{\hbar^2}{2m}
\left(\frac{3\pi^2}{2}\right)^{2/3}\rho^{5/3}F_{5/3} \nonumber \\
&+&\frac{1}{8}t_0\rho^2\left[2(x_0+2)-(2x_0+1)F_2\right] \nonumber\\
&+&\frac{1}{48}t_3\rho^{(\sigma+2)}\left[2(x_3+2)-(2x_3+1)F_2\right]\nonumber\\
&+&\frac{3}{40}(\frac{3\pi^2}{2})^{2/3} \rho^{8/3}
\left[t_1(x_1+2)+t_2(x_2+2)\right]F_{8/3} \nonumber \\
&+&\frac{3}{8}(\frac{3\pi^2}{2})^{2/3} \rho^{8/3}
\left[t_2(2x_2+1)-t_1(2x_1+1)\right]F_{8/3}. \nonumber
\label{eq1}
\end{eqnarray}
The symmetry energy $E_{sym} (\rho)$, slope parameter $L (\rho)$,
symmetry incompressibility $K_{sym} (\rho)$ and incompressibility
at saturation
$K_0 (\rho_0)$ can be derived from the energy density, which are explicitly
given in Refs. \cite{dutra12}.
The 13 Skyrme parameter sets used in the present calculations are 
SGII \cite{sgii}, SkM* \cite{skmstar},  RATP \cite{ratp},
SLy23a \cite{sly23ab}, SLy23b \cite{sly23ab}, SLy4 \cite{sly45}, 
SLy5 \cite{sly45}, SkT1 \cite{skt}, SkT2 \cite{skt}, KDE0v1 \cite{kde0v1},
LNS \cite{lns}, NRAPR \cite{nrapr}, SkMP \cite{skmp} and displayed in Table I.


\subsection{Relativistic mean field (RMF) formalism}

In principle, one should use quantum chromodyanmics (QCD), the
fundamental theory of strong interaction, for the complete
description of EOS. But it cannot be use to describe hadronic matter
due to its non-perturbative properties. A major breakthrough
occurred when the concept of effective field theory (EFT) was
introduced and applied to low energy QCD \cite{we79}. 
The degrees of freedom in this theory
are nucleons interacting through the exchange of iso-scalar scalar
$\sigma$, iso-scalar vector $\omega$, iso-vector-vector $\rho$ and
the pseudoscalar $\pi$ mesons. The nucleons are considered as Dirac
particle moving in classical meson fields. The contribution of $\pi$
meson is zero at mean field level, due to pseudo-spin nature. The
chiral effective Lagrangian (E-RMF) proposed by Furnstahl, Serot and Tang
\cite{fu96,g,mu96} is the extension of the standard relativistic
mean field (RMF) theory \cite{l1,bo77} with the addition of non-linear
scalar-vector and vector-vector self interaction. This Lagrangian
includes all the non-renormalizable couplings consistent with the
underlying symmetries of QCD. Applying the naive dimensional
analysis \cite{ge84,ge93} and the concept of naturalness one can
expand the nonlinear Lagrangian and organize it in increasing powers
of the fields and their derivatives and truncated at given level of
accuracy \cite{ru97,fu00,se04}. In practice, to get a reasonable
result, one needs the Lagrangian up to 4th order of interaction. 
Thus, the considered model involves the nucleons interacting
through the mesons. The truncated Lagrangian which includes the terms up
to the fourth order is given by
\begin{eqnarray}
{\cal L} & = & \overline{\Psi}_{B}\left ( i\gamma^\mu
D_{\mu} - m_{B} + g_{\sigma B}{\sigma}\right)
{\Psi}_{B}
+\frac{1}{2}{\partial_{\mu}}{\sigma}{\partial^{\mu}}{\sigma} \nonumber \\
&&-m_{\sigma}^2{\sigma^2}\left(\frac{1}{2}
+\frac{\kappa_3}{3 !}
\frac{g_{\sigma B}\sigma}{m_{B}}+\frac{\kappa_4}{4 !}
\frac{g^2_{\sigma B}\sigma^2}{m_{B}^2}\right)
-\frac{1}{4} {\Omega_{\mu\nu}}{\Omega^{\mu\nu}}
\nonumber \\
& & +\frac{1}{2}\left (1 +
{\eta_1}\frac{g_{\sigma B}\sigma}{m_{B}}
+\frac{\eta_2}{2}\frac{g^2_{\sigma B}\sigma^2}{m_{B}^2}\right)
m_{\omega}^2{\omega_{\mu}}
{\omega^{\mu}}
\nonumber \\
&&- \frac{1}{4}{R^a_{\mu\nu}}{R^{a\mu\nu}}
+\left(1 + \eta_{\rho}\frac{g_{\sigma B}\sigma}{m_{B}}\right)
\frac{1}{2}m_{\rho}^2{\rho^a_{\mu}}{\rho^{a\mu}}
\nonumber \\
&&+\frac{1}{4 !}{\zeta_{0}}
g^2_{\omega B}\left({\omega_{\mu}}{\omega^{\mu}}\right)^2.
\label{lag}
\end{eqnarray}
The subscript
$B=n,p$ denotes for nucleons. The terms in
eqn. (\ref{lag}) with the subscript $B$ should be interpreted as sum over
the states of nucleons. The covariant derivative ${D_{\mu}}$ is defined as
\begin{eqnarray}
{D_{\mu}} & = & \partial_{\mu} + ig_{\omega B}{\omega_{\mu}}
+ ig_{\phi B}{\phi_{\mu}} + ig_{\rho B}I_{3B}{\tau^a}{\rho^a_{\mu}},
\end{eqnarray}
whereas $R^a_{\mu\nu}$, and $\Omega_{\mu\nu}$ are the field tensors
\begin{equation}
R^a_{\mu\nu}= \partial_{\mu}\rho^a_{\nu} -
\partial_{\nu}\rho^a_{\mu} +
g_{\rho}\epsilon_{abc}\rho^b_{\mu}\rho^c_{\nu},
\end{equation}
\begin{equation}
{\Omega_{\mu\nu}} =  \partial_{\mu}\omega_{\nu} -
\partial_{\nu}\omega_{\mu},
\end{equation}
where $m_{B}$ denotes the baryon and $m_\sigma$, $m_\omega$,
$m_\rho$ are the masses assigned to the meson fields. Using this
Lagrangian, we derive the equation of motion and solved it in the
mean field approximation self consistently. Here,
the meson fields are replaced by their classical expectation
values. The field equations for $\sigma$, $\omega$ and $\rho$-meson
are given by
\begin{widetext}
\begin{eqnarray}
m^2_\sigma \left(\sigma_0 + \frac{g_{\sigma
B}\kappa_{3}}{2m_{B}}{\sigma^2_0} +\frac{g^2_{\sigma
B}\kappa_{4}}{6m_{B}^2}{\sigma^3_0}\right)  
-\frac{1}{2}m^2_{\rho}\eta_{\rho}\frac{g_{\sigma}}{m_{B}}\rho^2_0 
-\frac{1}{2}m^2_{\omega}
\left(\eta_1\frac{g_{\sigma B}}{m_{B}} + \eta_2\frac{g^2_{\sigma
B}}{m_{B}^2}\sigma_0\right) {\omega^2} 
= \sum_{B}g_{\sigma B}\rho_{SB},
\end{eqnarray}

\begin{eqnarray}
& & m^2_{\omega}\left(1 + \frac{{\eta_1}g_{\sigma}}{m_{B}}\sigma_0 +
\frac{{\eta_2}g^2_{\sigma}}{2m_{B}^2}\sigma^2_0\right){\omega}_0 +
\frac{1}{6}{\zeta_0} g^2_{\omega B}{\omega^3_0} 
= \sum_{B}g_{\omega B}\rho_B,
\end{eqnarray}
\end{widetext}
and 
\begin{eqnarray}
& & m^2_{\rho}\left(1 + \frac{g_{\sigma
B}{\eta_{\rho}}}{m_{B}}\sigma_0\right)\rho_{03} = \sum_{B}g_{\rho
B}I_{3B}\rho_B.
\end{eqnarray}
For a baryon species, the scalar density, $\rho_{SB}$, and baryon
density $(\rho_B)$ are
\begin{equation}
\rho_{SB}= \frac{2J_{B}+1}{2\pi^2}\int_{0}^{k_B} \frac{k^2
dk}{E^*_B}
\end{equation}
and 
\begin{equation}
\rho_{B} = \frac{2J_{B}+1}{2\pi^2}\int_{0}^{k_B} {k^2 dk},
\end{equation}
where $E^{*}_{B}=\sqrt{k^{2}_B+{m^*}^2_{B}}$ is the effective energy
and $J_{B}$ and $I_{3B}$ are the spin and isospin projection of
baryon $B$, the quantity $k_B$ is the Fermi momentum for the baryon,
$m^*=m_B-g_{\sigma B}\sigma$ is the effective mass, which is solved
self-consistently.
After obtaining the
self-consistent fields, the pressure ${\cal P}$ and total energy density
$\varepsilon $ for a given baryon density are
\begin{widetext}
\begin{eqnarray}
\cal P &=& \frac{\gamma}{3(2\pi )^{3}}\int_{0}^{k_B}
d^{3}k\frac{k^{2}}{E^{*}_{B}(k)}+\frac{1}{ 4!}\zeta _{0}g_{\omega
B}^{2}{\omega}_{0}^{4}+\frac{1}{2}\Bigg(1+\eta _{1}\frac{g_{\sigma
B} \sigma_{0}}{m_{B}}+\frac{\eta _{2}}{2}\frac{g_{\sigma
B}^{2}\sigma_{0}^{2}}{m_{B}^{2}}\Bigg)
m_{\omega B}^{2}{\omega}_{0}^{2}  \nonumber \\
&&\null -m_{\sigma
B}^{2}\sigma_{0}^{2}\Bigg(\frac{1}{2}+\frac{\kappa _{3}g_{\sigma
B}\sigma _{0}}{3!m_{B}}+\frac{\kappa _{4}g_{\sigma
B}^{2}\sigma_{0}^{2}}{4!m_{B}^{2}}\Bigg)+\frac{1}{2 }\Bigg(1+\eta
_{\rho }\frac{g_{\sigma B}\sigma_{0}}{m_{B}}\Bigg)m_{\rho
}^{2}\rho_{0}^{2} + {P}_l,
\label{eqFN25}
\end{eqnarray}
and 
\begin{eqnarray}
\cal E &=& \frac{\gamma}{(2\pi )^{3}}\int_{0}^{k_B} d^{3}k
E^{*}_{B}(k)+\frac{1}{ 8}\zeta _{0}g_{\omega
B}^{2}{\omega}_{0}^{4}+\frac{1}{2}\Bigg(1+\eta _{1}\frac{g_{\sigma
B} \sigma_{0}}{m_{B}}+\frac{\eta _{2}}{2}\frac{g_{\sigma
B}^{2}\sigma_{0}^{2}}{m_{B}^{2}}\Bigg)
m_{\omega B}^{2}{\omega}_{0}^{2}  \nonumber \\
&&\null +m_{\sigma B}^{2}\sigma_{0}^{2}\Bigg(\frac{1}{2}+
\frac{\kappa _{3}g_{\sigma B}\sigma_{0}}{3!m_{B}}+\frac{\kappa
_{4}g_{\sigma B}^{2}\sigma _{0}^{2}}{4!m_{B}^{2}}\Bigg)+\frac{1}{2
}\Bigg(1+\eta _{\rho }\frac{g_{\sigma B}\sigma
_{0}}{m_{B}}\Bigg)m_{\rho }^{2}\rho_{0}^{2} + {\varepsilon}_l,
\end{eqnarray}
\end{widetext}
here, $\gamma$ is the spin degeneracy ($\gamma$=2 for pure neutron matter
and $\gamma$=4 for symmetric nuclear matter), 
$P_{l}$ and $\varepsilon_{l}$ are lepton pressure and energy density,
respectively. 
For the stability of neutron star in which the strongly interacting 
particles are baryons,
the composition is determined by the requirements of charge
neutrality and $\beta$-equilibrium conditions under the weak
processes $B_1 \to B_2 + l + {\overline \nu}_l$ and $B_2 + l \to B_1
+ \nu_l$. After deleptonization, the charge neutrality condition
yields
\begin{equation}
q_{\rm tot} = \sum_B q_B (2J_B + 1) k_B^3 \big/ (6\pi^2) +
\sum_{l=e,\mu} q_l k_l^3 \big/ (3\pi^2)  = 0 ~, \label{neutral}
\end{equation}
where $q_B$ corresponds to the electric charge of baryon species $B$
and $q_l$ corresponds to the electric charge of lepton species $l$.
Since the time scale of a star is effectively infinite compared to
the weak interaction time scale, weak interaction violates
strangeness conservation. The strangeness quantum number is
therefore not conserved in a star and the net strangeness is
determined by the condition of $\beta$-equilibrium, which for baryon
$B$ is given by $\mu_B = b_B\mu_n - q_B\mu_e$, where $\mu_B$ is
the chemical potential of baryon $B$ and $b_B$ its baryon number.
Thus the chemical potential of any baryon can be obtained from the
two independent chemical potentials $\mu_n$ and $\mu_e$ for neutron
and electron, respectively.
The lepton Fermi momenta are the positive real solutions of $(k_e^2
+ m_e^2)^{1/2} =  \mu_e$ and $(k_\mu^2 + m_\mu^2)^{1/2} = \mu_\mu =
\mu_e$. The equilibrium composition of the star is obtained by
putting the $\beta$- equilibrium with the charge
neutrality condition  Eqn. (\ref{neutral}) at a given total baryonic
density $\rho = \sum_B (2J_B + 1) k_B^3/(6\pi^2)$; the baryon
effective masses are obtained self-consistently.
In our calculation, we have taken 7 well established parameter sets 
such as:  
G2 \cite{g}, G1 \cite{g}, NL3 \cite{nl3}, TM1 \cite{tm1}, FSU \cite{fsu}, 
L1 \cite{l1}, SH \cite{sh}. These all parameters along with their 
saturation properties are given in Table \ref{table1}.  


\subsection{Stellar Equations}
In the interior part of neutron star, the neutron chemical potential exceeds
the combined masses  of the proton and electron. Therefore, asymmetric matter
with an admixture of electrons rather than pure neutron matter, is a more
likely composition of matter in neutron star interiors. The concentrations
of neutrons, protons and electrons can be determined from the condition of
$\beta-$equilibrium
$n\leftrightarrow p+e+{\bar{\nu}}$
and from charge neutrality, assuming that neutrinos are not degenerate. 
Here n, p, $\nu$ have their usual meaning of neutron, proton and neutrino, 
respectively.
In momentum conservation condition
$\nu_n=\nu_p+\nu_e,$ $n_p=n_e$, where
$\nu_n=\mu_n-g_{\omega}V_0 + \frac{1}{2}g_{\rho}b_0$ and
$\nu_p=\mu_p-g_{\omega}V_0 - \frac{1}{2} g_{\rho}b_0$ with
$\mu_n={\sqrt{({k_{fn}^2}+{M^*{^2}_n})}}$ and
$\mu_p={\sqrt{({k_{fp}^2}+{M^*{^2_p} })}}$ 
are the chemical potential, and $k_{fn}$ and $k_{fp}$ are the Fermi
momentum for neutron and proton, respectively.
Imposing this conditions, in the expressions of ${\cal E}$ and ${\cal P}$, 
we evaluate ${\cal E}$ and ${\cal P}$ as a function of density.
To calculate the star structure, we use
the Tolman-Oppenheimer-Volkoff (TOV) equations for the structure of a
relativistic spherical and
static star composed of a perfect fluid were derived from Einstein's
equations \cite{tov}, where the pressure and energy densities 
are only the input ingredients.
The TOV equation is given by  \cite{tov}:

\begin{equation}\label{tov1}
\frac{d{\cal P}}{dr}=-\frac{G}{r}\frac{\left[{\cal E} + \cal P\right ]
\left[M+4\pi r^3 \cal P \right ]}{(r-2 GM)},
\end{equation}

\begin{equation}\label{tov2}
\frac{dM}{dr}= 4\pi r^2 \cal E,
\end{equation}
with $G$ as the gravitational constant and $M(r)$ as the enclosed
gravitational mass. We have used $c=1$, the velocity of light. 
Given the ${\cal P}$ and ${\cal E}$, these equations can be integrated
from the origin as an initial
value problem for a given choice of central energy density.
The value of $r~(=R)$, where the pressure vanishes defines the
surface of the star.
Another realistic approximation that when neutron star is rotating with static,
axial symmetric, space-time, the time translational invariant and 
axial-rotational invariant metric in spherical polar coordinate 
(t, r, $\theta$, $\phi$) can be written as:

\begin{eqnarray}
ds^2 &=& -e^{2\nu} dt^{2} + e^{2\alpha}(dr^{2} + r^2 d\theta ^2) \nonumber \\ 
&& + e^{2\beta} r^2 sin^2 \theta(d\phi - \omega dt)^2 , 
\label{ds2}
\end{eqnarray}
where the metric functions $\nu$, $\alpha$, $\beta$, $\omega$ depend 
only on r and $\theta$. For a perfect fluid, the energy momentum tensor
can be given by: 
\begin{equation}
T^{\mu \nu} = Pg^{\mu\nu} + ({\cal P + E}) u^{\mu}u^{\nu},
\label{t}
\end{equation}
with the four-velocity 
\begin{equation}
u^{\mu}=\frac{e^{-\nu}}{\sqrt{1-v^2}} (1, 0, 0, \Omega),
\label{u}
\end{equation}
here
\begin{equation}
v=(\Omega-\omega)r\; sin \;\theta e^{\beta-\nu},
\label{v}
\end{equation}
is the proper velocity relative to an observer with zero angular 
velocity and $\Omega$ is the angular velocity of the star measured from 
infinity. Now, we can compute the Einstein field equation given by
\begin{equation}
R_{\mu\nu} - \frac{1}{2} g_{\mu\nu} R = 8\pi T_{\mu\nu},
\label{r}
\end{equation}
where $R_{\mu \nu}$ is Ricci tensor and R is the scalar curvature. 
From this, we can solve the equation of motion for metric function:
\begin{eqnarray}
\Delta \left[ \rho e ^{\zeta} \right]&=& S_{\rho}(r,\mu) , \\ [1mm]
\left( \Delta + \frac{1}{r} \frac{\partial}{\partial r} 
-\frac{1}{r^2}\mu \frac{\partial}{\partial r}\right) \gamma e^{\zeta} 
&=& S_{\gamma}(r,\mu), \\ [1mm]
\left( \Delta + \frac{2}{r} \frac{\partial}{\partial r} 
-\frac{2}{r^2}\mu \frac{\partial}{\partial r}\right) 
\omega e^{\frac{\gamma - 2\rho}{2}}  
&=& S_{\omega}(r,\mu),  
\end{eqnarray}
where $\gamma=\beta+v$, $\rho=v-\beta$ and $\mu=$cos$ \theta$. The right hand 
side of equations are the source terms. One can find more details about 
these equations in Ref. \cite{komatsu89}.
We can put the limit on the maximum rotation i.e. Kepler frequency $\Omega_k$, 
by the onset of mass shedding from equator of the star. 
The final expression for $\Omega_k$, in general relativistic formalism 
is given as:
\begin{equation}
\Omega_{K} = \omega + \frac{\omega'}{2 \psi'} +
e^{v-\beta} \left[ \frac{1}{R^2}\frac{v'}{\psi'} + 
\left(\frac{e^{\beta-v} \omega'}{2\psi'} \right)^2 \right]^{\frac{1}{2}},
\end{equation}
where $\psi = \beta' + \frac{1}{R}$ and the prime denotes the differentiation
with respect to the radial coordinate.
For the calculation of 
rotational neutron star properties like mass, radius, rotational frequency,
we used the well established rotational neutron star (RNS) code, 
which is written by Stergioulas \cite{gle94, ster95}.

\subsection{Properties of Rotating Neutron Star}
We have calculated the maximum mass and radius of static and
rotating neutron star by using well established RNS code. For this, we need only
energy and pressure density which will be provided by non-relativistic
and relativistic models of equation of state. 
Now, our aim is to calculate maximum $m=2$
quadrupole moment for neutron star by using a chemically detailed model
for the crust \cite{benjamin05}. The relation of quadrupole moment  with
maximum mass $M$ and  radius $R$ is given as:

\begin{eqnarray}
\Phi_{22}&=&2.4\times10^{38} g {cm}^2 \left(\frac{\sigma_{max}}{10^{-2}}\right)
\left(\frac{R}{10 km}\right)^{6.26} \nonumber \\
&& \times \left(\frac{1.4M_\odot}{M}\right)^{1.2} ,
\label{quadra} 
\end{eqnarray}
where $\sigma_{max}$ is called breaking strain of the crust. In our calculation
we have taken its two possible values i.e. $10^{-2},\; 10^{-3}$.

The quadrupole moment [Eqn. (\ref{quadra})] and ellipticity of the neutron
star is connected to each other by a simple relation~\cite{benjamin05}:
\begin{eqnarray}
\label{eps} 
\epsilon=\sqrt{\frac{8\pi}{15}} \frac{\Phi_{22}}{I_{zz}},
\end{eqnarray}
where the z axis is the rotation axis and $I_{zz}$ is the moment of inertia 
along the z-axis and for conventional neutron star, it is 
given as \cite{bej02}:
\begin{eqnarray}
\label{moment}
I_{zz}&=&9.2\times 10^{44} g cm^2 \left(\frac{M}{1.4M_\odot}\right)\left( 
\frac{R}{10 km}\right)^2 \nonumber \\
&&\times \left[1+0.7\left(\frac{M}{1.4M_\odot}\right)
\left(\frac{10km}{R}\right)\right].
\end{eqnarray}

For each (non-relativistic and relativistic) parameter set we can calculate the 
maximum mass and radius of the neutron star and then other observables like
quadrupole ellipticity and moment of inertia. The maximum rotational frequency 
$\nu_{max}$ of the stable rotationary neutron star can be given by the simple
relation \cite{sly23ab}.  

\begin{eqnarray} 
\label{frequency}
\nu_{max}=1.22\times10^{3}\left(\frac{M}{M_\odot}\right)^{1/2}
\left(\frac{R}{10 km}\right)^{-3/2},
\end{eqnarray}

Finally, we use eqns. (\ref{quadra} -
\ref{frequency}) to calculate the 
gravitational wave strain amplitude which is presented by \cite{prd07}:

\begin{eqnarray} 
\label{h_0} 
h_0=\frac{16 \pi^2 G}{c^4}\frac{\epsilon I_{zz} \nu^2}{r},
\end{eqnarray}
where $r$ is the distance of neutron star from the earth~\cite{jara98}.

\section{Results and discussions}

In this work, we have taken conventional static and rotating neutron star and 
perform the calculation for their mass and radius by using the TOV and 
RNS equations. The ingredients require to solve these two equations are 
pressure and energy density. 
After getting the mass and radius, we have calculated the other properties 
like quadrupole moment, ellipticity, moment of inertia and gravitational 
wave amplitude of rotating neutron star. 
We took the recently reported maximum mass and radius of neutron star
pulsar J1614-2230 \cite{demorest10} and some theoretical Dirac-Bruckner
Hartree-Fock results as a reference, where the star mass is 
$(1.97 \pm 0.04)M_\odot$. This means that an equation of state can be 
appreciated, if it has the capability to estimate a maximum mass at least 
$2.0 M_\odot$.


\begin{table*}
\caption{The Skyrme and RMF force parameters and their nuclear matter 
properties, like 
BE/A (MeV), compressibility $K_0$ (MeV), nucleon effective mass
ratio M*/M, symmetry energy $E_{sym}$ (MeV), $L_{sym}$ (MeV),
$K_{sym}$ (MeV) at saturation density $\rho_0$.
}
\begin{tabular}{|c|c|c|c|c|c|c|c|c|c|c|c|c|c|c|c|c|c|}
\hline
\multicolumn{18}{|c|}{Skyrme effective interaction}\\
\hline
\multicolumn{1}{|c|}{ } & \multicolumn{9}{|c|}{Coupling Constants } & \multicolumn{8}{|c|} { Nuclear Saturation Properties} \\
\cline{2-18}
Parameter              & $t_0$     & $t_1$  & $t_2$  & $t_3$    & $x_0$  & $x_1$ 
  & $x_2$ & $x_3$  & $\sigma$ & $\rho_0$  & $M^*/M$     & $BE/A$        & $K_0$       
& $E_{sym}$      & $L_{sym}$ & $K_{sym}$ & \\
\hline
SGII    \cite{sgii}  	&	-2645.0	&	340.0	&	-41.9	&	15595.0	&	0.09	&	-0.06	&	1.43	&	0.06	&	0.17	&	0.16	&	0.79	&	-15.6	&	214.7	&	26.8	&	37.6	&	-145.9  &	\\
SkM*   \cite{skmstar}	&	-2645.0	&	410.0	&	-135.0	&	15595.0	&	0.09	&	0.00	&	0.00	&	0.00	&	0.17	&	0.16	&	0.79	&	-15.8	&	216.6	&	30.0	&	45.8	&	-155.9	  &\\
SkMP     \cite{skmp} 	&	-2372.2	&	503.6	&	57.3	&	12585.3	&	-0.16	&	-0.40	&	-2.96	&	-0.27	&	0.17	&	0.16	&	0.65	&	-15.6	&	230.9	&	29.9	&	70.3	&	-49.8  &	\\
RATP    \cite{ratp}  	&	-2160.0	&	513.0	&	121.0	&	11600.0	&	0.42	&	-0.36	&	-2.29	&	0.59	&	0.20	&	0.16	&	0.67	&	-16.1	&	239.5	&	29.3	&	32.4	&	-191.2  &	\\
SLy23a \cite{sly23ab}	&	-2490.2	&	489.5	&	-566.6	&	13803.0	&	1.13	&	-0.84	&	-1.00	&	1.92	&	0.17	&	0.16	&	0.70	&	-16.0	&	229.9	&	32.0	&	44.3	&	-98.2  &	\\
SLy23b \cite{sly23ab}	&	-2488.9	&	486.8	&	-546.4	&	13777.0	&	0.83	&	-0.34	&	-1.00	&	1.35	&	0.17	&	0.16	&	0.69	&	-16.0	&	229.9	&	32.0	&	46.0	&	-119.7  &	\\
SLy4   \cite{sly45}  	&	-2488.9	&	486.8	&	-546.4	&	13777.0	&	0.83	&	-0.34	&	-1.00	&	1.35	&	0.17	&	0.16	&	0.69	&	-16.0	&	229.9	&	32.0	&	45.9	&	-119.7  &	\\
SLy5   \cite{sly45}  	&	-2483.5	&	484.2	&	-556.7	&	13757.0	&	0.78	&	-0.32	&	-1.00	&	1.26	&	0.17	&	0.16	&	0.70	&	-16.0	&	229.9	&	32.0	&	48.2	&	-112.8  &	\\
SkT1   \cite{skt}    	&	-1794.0	&	298.0	&	-298.0	&	12812.6	&	0.15	&	-0.50	&	-0.50	&	0.09	&	0.33	&	0.16	&	1.00	&	-16.0	&	236.2	&	32.0	&	56.2	&	-134.8  &	\\
SkT2   \cite{skt}    	&	-1791.6	&	300.0	&	-300.0	&	12792.0	&	0.15	&	-0.50	&	-0.50	&	0.09	&	0.33	&	0.16	&	1.00	&	-15.9	&	235.7	&	32.0	&	56.2	&	-134.7  &	\\
KDE0v1 \cite{kde0v1} 	&	-2553.1	&	411.7	&	-419.9	&	14603.6	&	0.65	&	-0.35	&	-0.93	&	0.95	&	0.17	&	0.17	&	0.74	&	-16.2	&	227.5	&	34.6	&	54.7	&	-127.1  &	\\
LNS    \cite{lns}    	&	-2485.0	&	266.7	&	-337.1	&	14588.2	&	0.06	&	0.66	&	-0.96	&	-0.03	&	0.17	&	0.18	&	0.83	&	-15.3	&	210.8	&	33.4	&	61.5	&	-127.4  &	\\
NRAPR  \cite{nrapr}  	&	-2719.7	&	417.6	&	-66.7	&	15042.0	&	0.16	&	-0.05	&	0.03	&	0.14	&	0.14	&	0.16	&	0.69	&	-15.9	&	225.7	&	32.8	&	59.6	&	-123.3  &	\\
\hline
\hline
\multicolumn{18}{|c|}{Relativistic Mean field interaction}\\
\hline
\multicolumn{1}{|c|}{ } & \multicolumn{10}{|c|}{Coupling Constants } & \multicolumn{7}{|c|} {Nuclear Saturation Properties} \\
\cline{2-18}
Parameter & $g_\sigma$         & $g_\omega$ & $g_\rho$ & $k3$   & $k4$    
& $\zeta_0$ & $\eta_1$ & $\eta_2$ & $\eta_r$ & $\Lambda_v $ & $\rho_0 $  
& $M^*/M$     & $BE/A$        & $K_0$       & $E_{sym}$      & $L_{sym}$ 
& $K_{sym}$ \\
\hline
 G2 \cite{g}   	&	0.84	&	1.02	&	0.76	&	3.25	&	0.63	&	2.64	&	0.65	&	0.11	&	0.39	&	0.00	&	0.15	&	0.66	&	-16.1	&	214.7	&	36.4	&	100.7	&	-7.4	\\
 G1 \cite{g}   	&	0.79	&	0.97	&	0.70	&	2.21	&	-10.09	&	3.53	&	0.07	&	-0.96	&	-0.27	&	0.00	&	0.15	&	0.60	&	-16.2	&	215.0	&	37.9	&	118.6	&	91.7	\\
 NL3 \cite{nl3}	&	0.81	&	1.02	&	0.71	&	1.47	&	-5.67	&	0.00	&	0.00	&	0.00	&	0.00	&	0.00	&	0.15	&	0.60	&	-16.3	&	271.8	&	37.4	&	118.9	&	103.4	\\
 TM1 \cite{tm1}	&	0.80	&	1.00	&	0.74	&	1.02	&	0.12	&	2.69	&	0.00	&	0.00	&	0.00	&	0.00	&	0.15	&	0.63	&	-16.3	&	281.1	&	36.9	&	110.6	&	33.8	\\
 FSU \cite{fsu}	&	0.84	&	1.14	&	0.94	&	0.62	&	9.75	&	12.27	&	0.00	&	0.00	&	0.00	&	0.03	&	0.15	&	0.61	&	-16.3	&	230.0	&	32.6	&	60.4	&	-50.5	\\
 L1 \cite{l1}  	&	0.76	&	0.93	&	0.00	&	0.00	&	0.00	&	0.00	&	0.00	&	0.00	&	0.00	&	0.00	&	0.19	&	0.56	&	-15.8	&	546.6	&	22.1	&	74.6	&	73.6	\\
 SH \cite{sh}  	&	0.83	&	1.10	&	0.64	&	0.00	&	0.00	&	0.00	&	0.00	&	0.00	&	0.00	&	0.00	&	0.15	&	0.54	&	-15.8	&	545.0	&	35.0	&	115.6	&	92.8	\\

\hline
\end{tabular}
\label{table1}
\end{table*}

\subsection{Equation of State}

\begin{figure}[h]
\includegraphics[width=1.\columnwidth]{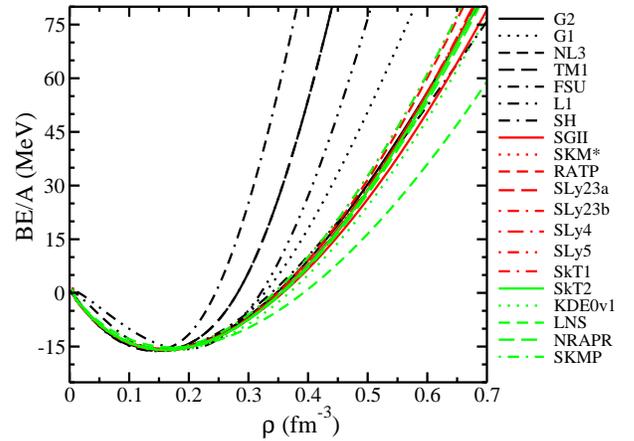}
\caption{ Binding energy per nucleon (MeV) for symmetric nuclear matter 
in for non-relativistic and relativistic models with 
baryon density }
\label{bea}
\end{figure}

\begin{figure}[h]
\includegraphics[width=1.\columnwidth]{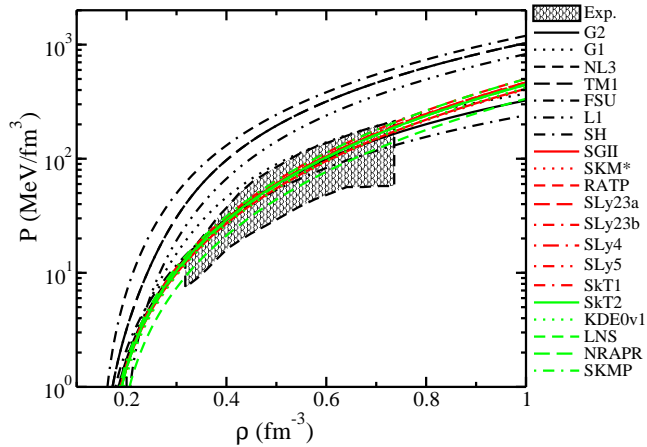}
\caption{Pressure density (MeV/fm$^{3}$) of symmetric nuclear matter 
non-relativistic and relativistic models with baryon density.  }
\label{pres}
\end{figure}

Most of the parameters are fitted to the saturation properties of 
symmetric nuclear matter like binding energy per nucleon ($BE/A$), 
effective mass of nucleons, incompressibility modulus $K_0$ 
and symmetry energy $E_{sym}$ at saturation density ($\rho_0$). 
We have shown these empirical 
values in Table \ref{table1} for both SHF and RMF parameter 
sets. For a general idea and to see the behaviour 
of these forces  
on binding energy per nucleon and pressure density, we have 
plotted figures (\ref{bea}) and (\ref{pres}). We get a stiff equation of 
state (EOS)
for SH parameter, which is one of the oldest RMF interaction 
and a soft EOS for LNS parameter, which is a successful set 
of SHF formalism. The rest of the EOS's for various parameter sets 
are between these two extremes.
Our theoretical EOS for RMF and SHF results are compared with the most 
accepted experimental data of Danielewicz et al. \cite{daniel} in 
Fig. \ref{pres}. From the figure, it is seen that all the EOS predicted 
with SHF formalism passes nicely through the experimental shaded region.
On the other hand, the RMF based EOS of 
NL3 \cite{nl3}, SH \cite{sh}, TM1 \cite{tm1} 
are far from the experimental observation.


\noindent
However, the recently proposed G1 and G2 sets of RMF formalism very much 
within the experimental shaded region. These parameters not only match 
with the EOS of Ref. \cite{daniel} but also predict the recent 
mass of neutron star \cite{demorest10}. 
In Fig. \ref{tov_mass}, we have shown the mass and radius trajectory of 
neutron star equation of state. Fig. \ref{tov_mass}(a) stands for 
$\frac{M_\odot}{M}$ verses central density and Fig. \ref{tov_mass}(b) 
is shown for $\frac{M_\odot}{M}$ as a function of neutron star radius 
for all the 20 
force parametrizations (see below).


\subsection{Mass and Radius of Neutron star}

\begin{figure}[h]
\includegraphics[width=1.\columnwidth]{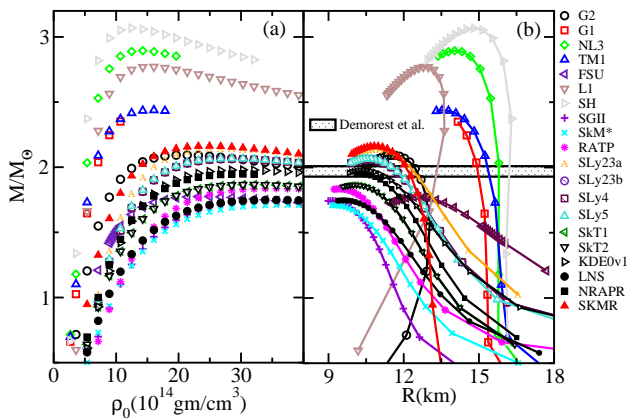}
\caption{Mass and radius trajectory for neutron star obtained from 
various parameter sets (equation of state) by using TOV equation. }
\label{tov_mass}
\end{figure}

\begin{figure}[h]
\includegraphics[width=1.\columnwidth]{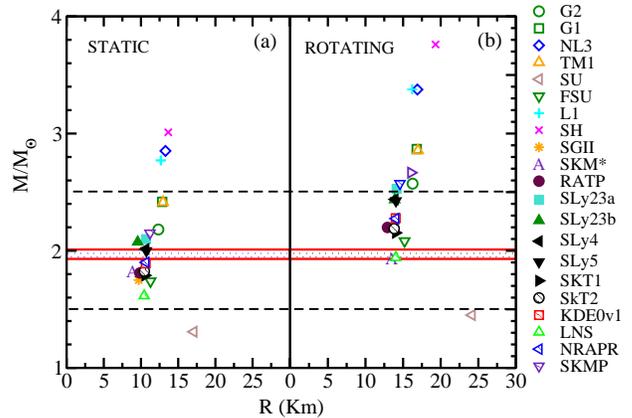}
\caption{Maximum mass ($M$/$M_\odot$) and radius $R$ (km) of static and 
rotating neutron star in RNS model with various non-relativistic and 
relativistic model parameters . }
\label{rns_mass}
\end{figure}

We noted down the maximum mass and the corresponding radius obtained from 
various non-relativistic and relativistic parameter sets [from TOV solution
Fig. \ref{tov_mass}]. Again from the RNS code, we collected the $M_{max}$ and 
$R_{max}$ for all sets. These masses and radii are depicted in Fig. 
\ref{rns_mass}(a) and Fig. \ref{rns_mass}(b) for static and rotating cases, 
respectively.
Here also, we put the maximum mass results of pulsar J1614-2230 
\cite{demorest10} as a standard reference and compared our results. 

If we compare the mass of static and rotational star, one can easily see 
that rotational neutron star
mass is larger compare to static one for the same parameter set. 
As Demorest et al. \cite{demorest10} stated that the theoretical models 
should have the 
maximum mass more or near to ($1.97\pm0.04)M_\odot$. The Shapiro delay
provides no information for the neutron star's radius, so we can not put 
any constraint on the radius of neutron star (NS).  
If we see the maximum mass and corresponding radius prediction of relativistic 
model parameter SU(2) (effective chiral model) \cite{su}, in both cases 
static and rotational, it is 
not suited to the current experimental observation \cite{demorest10}.
The reason is the extra softness of SU(2) model, that the vector meson
mass $(m_{\omega})$ is generated dynamically, as a results of which the 
effective mass of nucleon acquires a density dependence on both scalar 
and vector fields. The consequence of this dependency, the effective mass  
increases at higher density and EOS became more softer \cite{su}. 
Another non-relativistic model parameter LNS \cite{lns} which is not 
comfortable in static case, but 
it is within the cut off region for rotational NS.  

We compared our calculated 
results with experimental data of Demorest et al. \cite{demorest10} 
which is 
shown by the horizontal strip in figure \ref{tov_mass}(b). 
From this figure, a larger number of parameter sets,  
like FSU \cite{fsu}, SGII \cite{sgii}, SkM* \cite{skmstar}, 
LNS \cite{lns}, RATP \cite{ratp} and SkT2 \cite{skt} 
are not crossing the horizontal strip, which is the
experimental constraint on static slowrly rotating neutron star mass 
($\frac{M}{M_\odot}$)~\cite{demorest10}. So, these 
parameter sets are not acceptable whole heartedly in such high density 
scenario and need some discussions. \\
(i) As we have mentioned earlier, all the relativistic and non-relativistic 
parameter sets are constructed at the saturation and since these are 
effective parameters, there is no guarantee that the extrapolation of 
these forces are still valid at extremely high density, \\
(ii) Secondly, as it is in neutron star 
many of the SHF forces agree well with the 
recent EOS experimental data of heavy ion collision \cite{demorest10}, 
however these sets deviate when tested in the neutron star scenario. 
To reproduce the recent star mass \cite{demorest10} (as the masses do not 
lie within the experimental strip). 
With respect to this limit, IUFSU \cite{iufsu} is an extension of 
FSU \cite{fsu} lie within the experimental constraint.
For non-relativistic sets, the forces are chosen by taking into 
consideration their success in finite nuclei.
For more descriptive study, we refer the readers go through 
Ref.~\cite{dutra12}, where one will get 214 SHF parameter 
sets and their applications to various systems. 

\subsection{Rotational and Gravitational Wave Frequency and Amplitude}

\begin{figure}[h]
\includegraphics[width=1.\columnwidth]{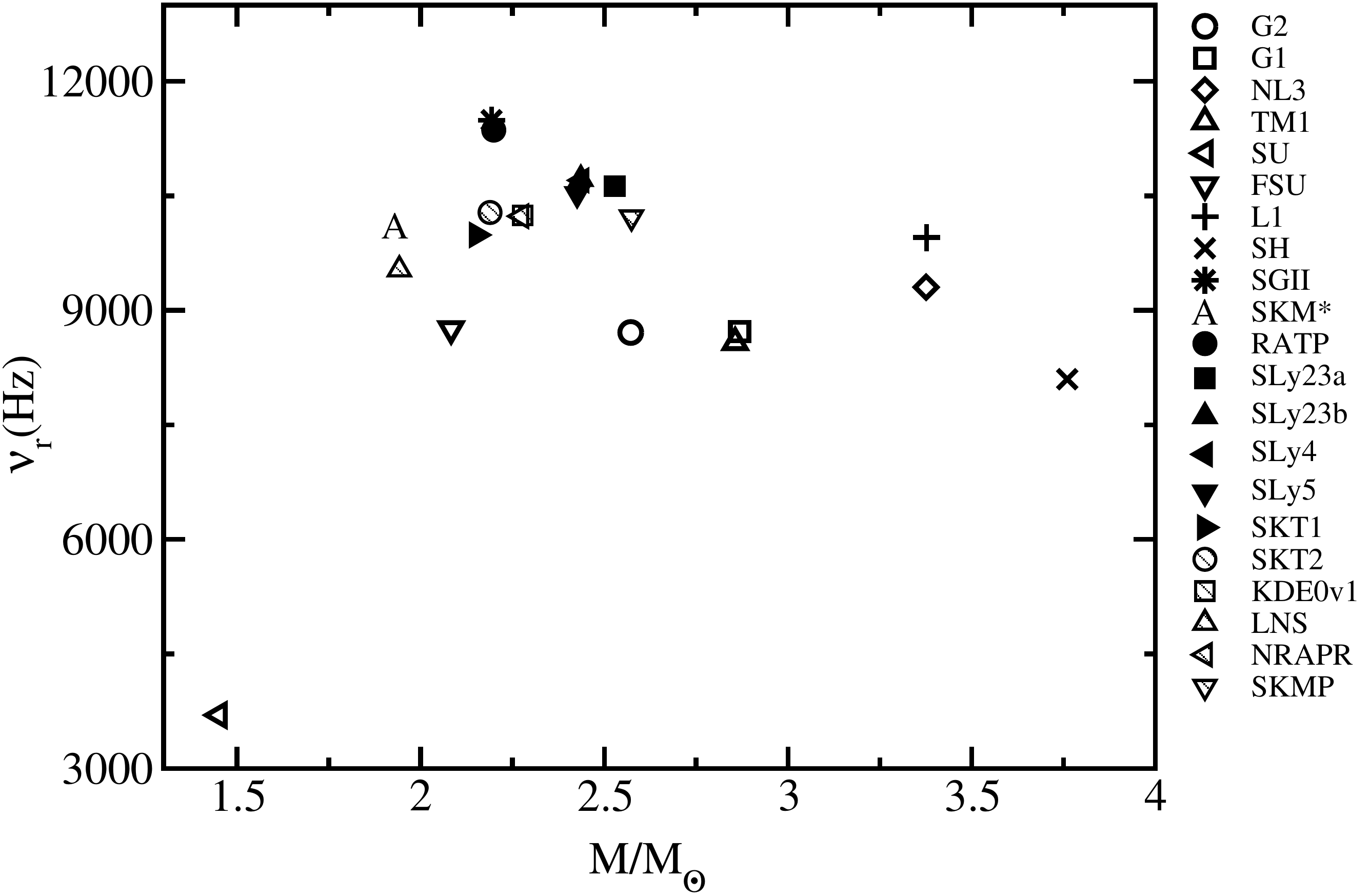}
\caption{ The rotational wave frequency ($\nu_{r}$) with maximum star mass
for various parameter sets. }
\label{mass_omega}
\end{figure}

\begin{figure}[h]
\includegraphics[width=1.\columnwidth]{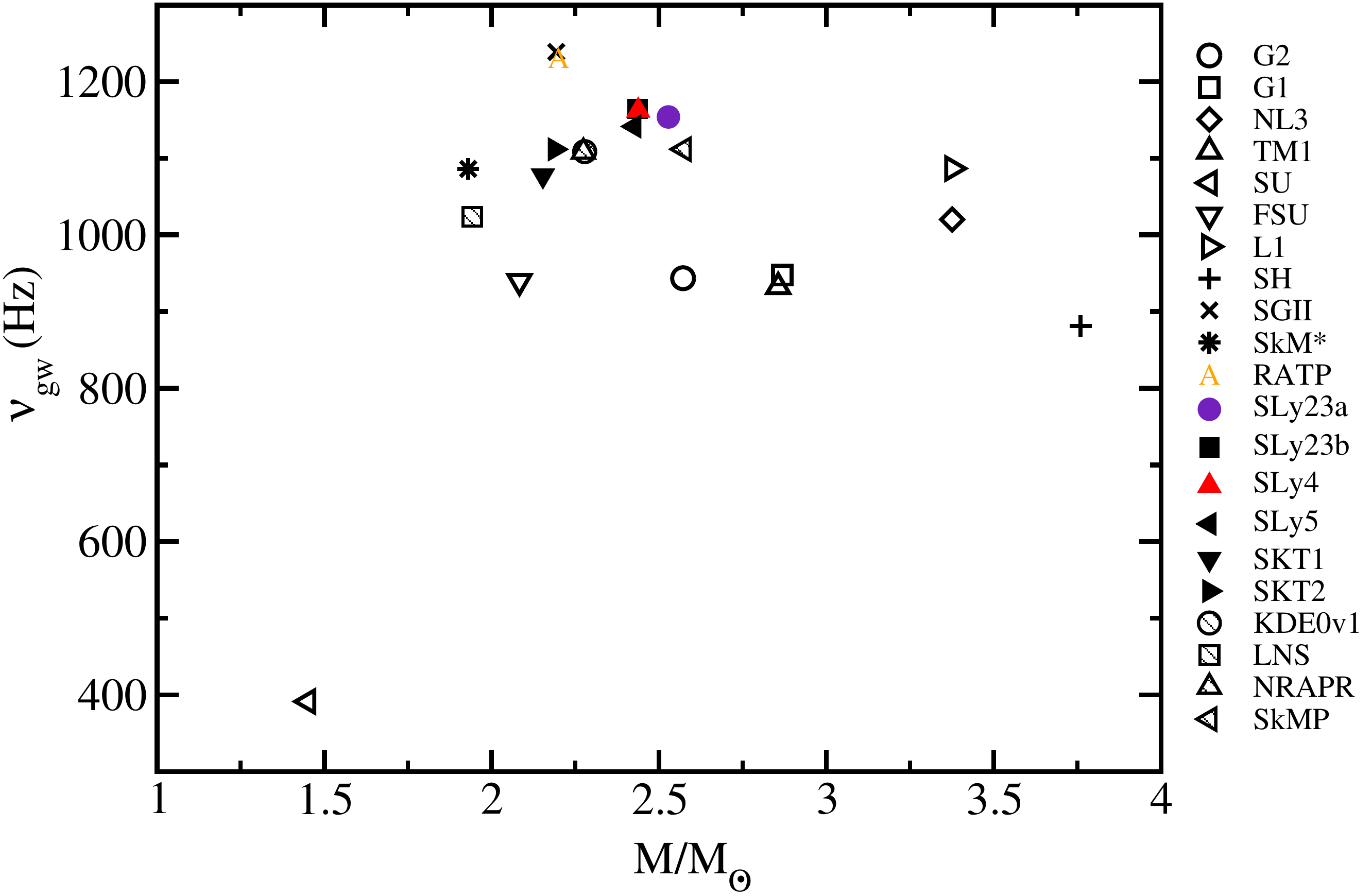}
\caption{ Gravitational wave frequency ($\nu_{gw}$) with maximum star mass
for various parameter sets. }
\label{mass_gw_freq}
\end{figure}

\begin{figure}[h]
\includegraphics[width=1.\columnwidth]{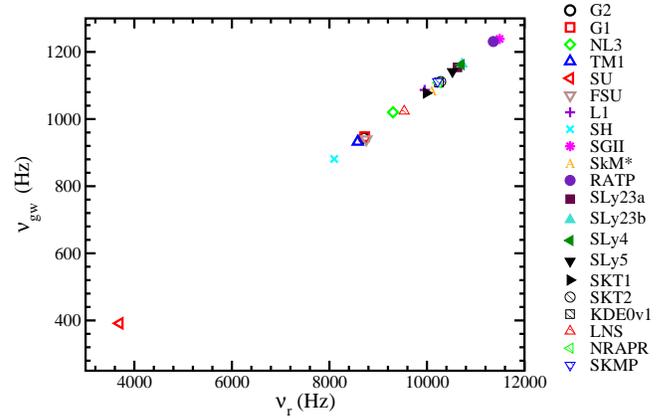}
\caption{ Correlation between rotational frequency of neutron star and 
emitted gravitational wave frequency in various parameter sets. }
\label{rot_gw_fre}
\end{figure}

Before going to discuss the gravitational wave frequency $\nu_{gw}$, we 
would like to see the rotational frequency $\nu_{r} = \frac{\Omega_K}{2\pi}$ 
of neutron star. The $\nu_{r}$ of a NS are found to be within 7000 to 12000 Hz 
(except SU(2) relativistic chiral parameter) for all the considered 
SHF and RMF parameter sets. The maximum rotational frequency 12000 Hz 
is predicted by the non-relativistic SGII \cite{sgii} and RATP \cite{ratp}
as shown in figure~\ref{mass_omega}. Unlike to the rotational Keplerian 
frequency $\Omega_K$, the gravitational frequency $\nu_{gw}$ is a tidious 
experimental exploration \cite{prd07,lisa}. The calculated values of
$\nu_{gw}$ obtained by various SHF and RMF parametrizations are shown in
figure~\ref{mass_gw_freq}. The value of $\nu_{gw}$ is found to be almost
9 times greater than $\nu_{r}$. A perfect coorelation between the gravitational
$\nu_{gw}$ and the rotational frequencies $\nu_{r}$ is shown in 
figure~\ref{rot_gw_fre}. From this figure, it is clear that increase of 
rotational results a larger emission of gravitational wave frequency 
$\nu_{gw}$.

\begin{figure}[h]
\includegraphics[width=1.\columnwidth]{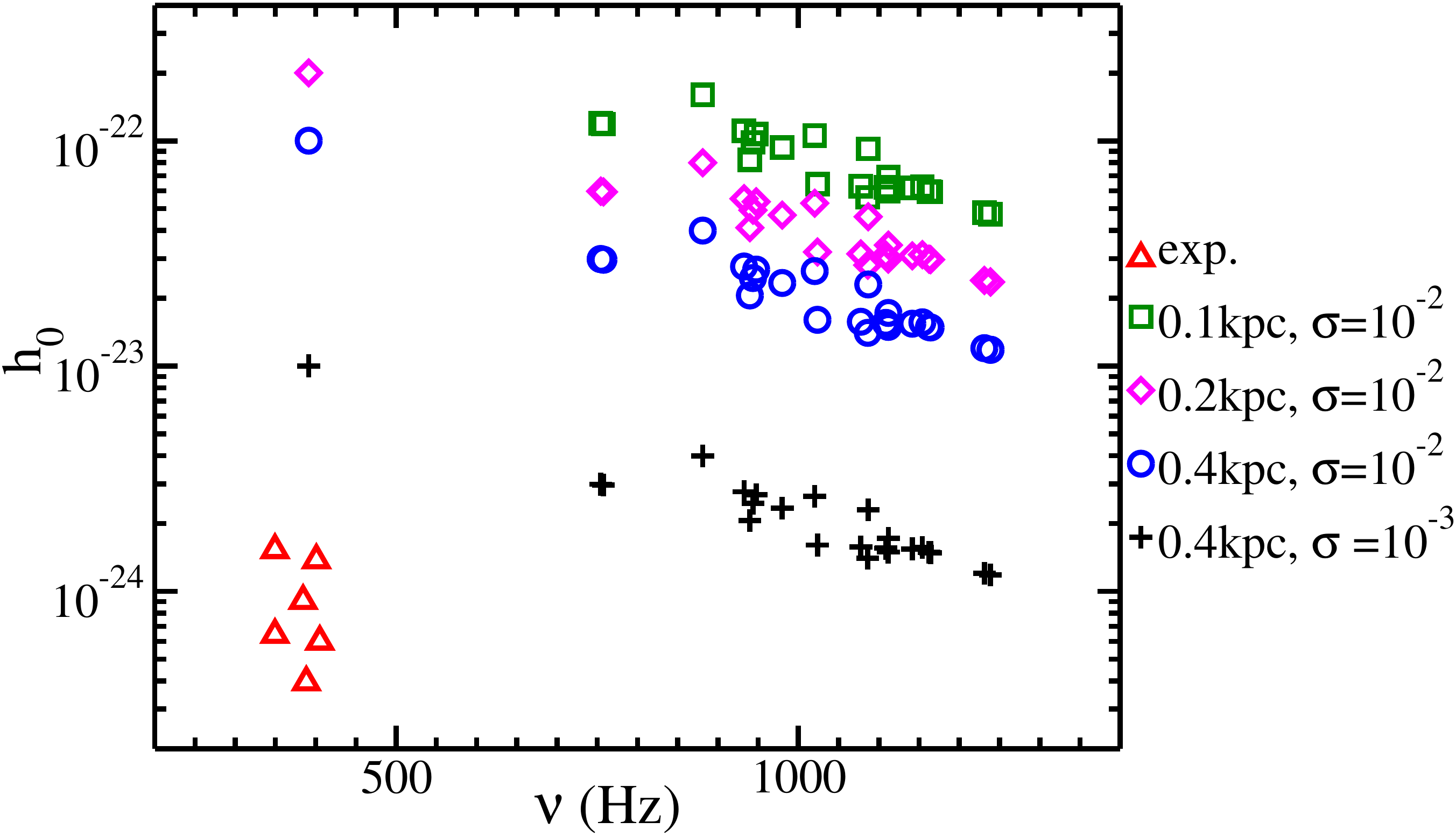}
\caption{Maximum gravitational wave (GW) strain amplitude $h_0$ with 
maximum possible GW frequency of rotational star.
}
\label{gw_fre}
\end{figure}

\noindent
For a rotating neutron star, the gravitational wave amplitude $h_0$ is an 
experimental observable. We can observed it directly by specially designed 
experimental setup \cite{prd07,lisa}. The gravitational wave is generated 
by the rotation of an axially asymmetric neutron star. 
The wave strain amplitude $h_0$ can be measured by knowing the maximum mass 
and corresponding radius of a star. Its analytical feeling can be 
taken from the equation (\ref{h_0}). The gravitational wave frequency 
$\nu_{gw}$ can be calculated by the equation (\ref{frequency}). The relation 
between gravitational wave strain $h_0$ amplitude and frequency $\nu_{gw}$
are shown in the figure~\ref{gw_fre}. 

In the calculations of 
quadrupole moment, we have taken two set of breaking strain of the neutron 
star crust $\sigma=10^{-2}, 10^{-3}$ and gravitational wave amplitude 
calculated with three sets of $r$ (0.1, 0.2 and 0.4 kpc) which is the distance 
between the star and earth. These are some standard values used by earlier 
calculations \cite{plamen08}. So in 
this way, we have given the GW strain amplitude and frequency relation for four 
set of data as shown in the figure \ref{gw_fre} along with the experimental 
results (for more discussion, see Ref. \cite{plamen08}).   
In our calculation, all gravitational frequencies come out more than $500$ Hz, 
except for SU(2) model ($\nu < 300 Hz)$. As we have mentioned earlier, this
parameter set is unable to produce maximum mass within the experimental 
limit \cite{demorest10}. 
We have noticed an important point here is that the gravitational wave 
strain amplitude decreases 
with increasing the $r$ and decreases with the value of breaking strain of 
neutron star crust $\sigma$.

\subsection{Quadrupole Moment of Neutron Star}

\begin{figure}[h]
\includegraphics[width=1.\columnwidth]{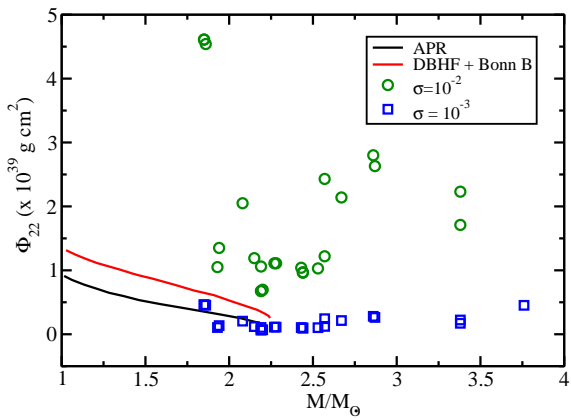}
\caption{Quadrupole deformation $\Phi_{22}$ with maximum mass 
of rotational neutron star }
\label{phi_mass}
\end{figure}

For quantitative understanding of the quadrupole moment $(\Phi_{22})$ 
in different relativistic and non-relativistic models parameters, 
we have calculated $(\Phi_{22})$ by using equation (\ref{quadra}). 
Although, it is not valid for high frequency rotating 
star, but for qualitative behaviour of model parameter, we can use this 
approximate relation, which depends only on the mass and radius of the 
neutron star with the breaking strain of the neutron star crust $\sigma$.
Presently, the $\sigma$ value is totally uncertain and its limiting ranges 
are $\sigma=(10^{-5}, 10^{-2})$ \cite{strain}. 
In the calculations, the two choosen values of 
$\sigma$ ($10^{-2}$ and $10^{-3}$) are taken to evaluate the quadrupole 
moments $\Phi_22$ and the results are shown in figure~\ref{phi_mass}. 
The results are also compared with the theoretical predictions of 
APR and DBHF + Bonn B.
The APR results shown by black line, which shows the variation of quadrupole 
moment of neutron star with mass, decreases continuously with $M$. 
Same trend we get in DBHF + Bonn B (red colour in Fig.~\ref{phi_mass}) 
predictions i.e. $\Phi_{22}$ with star mass. 
The results with $\sigma=10^{-3}$, match  well to the APR and DBHF + Bonn B 
predictions, while for $\sigma=10^{-2}$, we 
get very scattered values as shown in figure~\ref{phi_mass}.

\subsection{Moment of Inertia of Neutron Star}

\begin{figure}[h]
\includegraphics[width=1.\columnwidth]{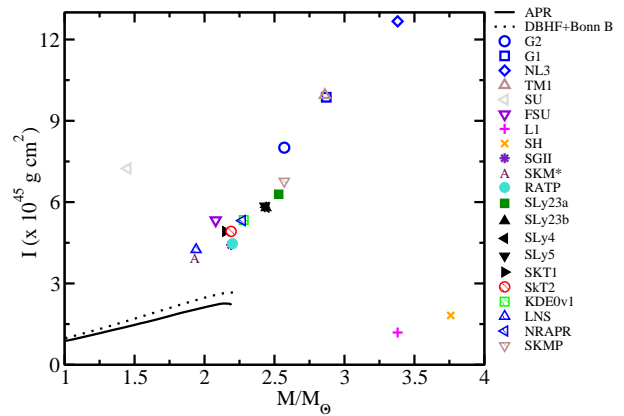}
\caption{ Moment of inertia (I) of rotational neutron star with mass at various 
parameter sets. }
\label{i_mass}
\end{figure}

In figure~\ref{i_mass}, we have given the moment of inertia (I) of rotating 
neutron star. Since, inertia is a static property, it is 
totally depend on its mass distribution. As we know from earlier discussion 
in this paper, the mass of the neutron star increases with the rotational 
frequency $\nu_r$. 
For calculating the moments of inertia (I) of the NS, we have used the 
maximum mass and corresponding radius and the obtained results are shown 
in the figure~\ref{i_mass}. The APR and DBHF + Born B results are also 
given in the figure for comparison. 

\subsection{Ellipticity of Neutron Star}

\begin{figure}[h]
\includegraphics[width=1.3\columnwidth]{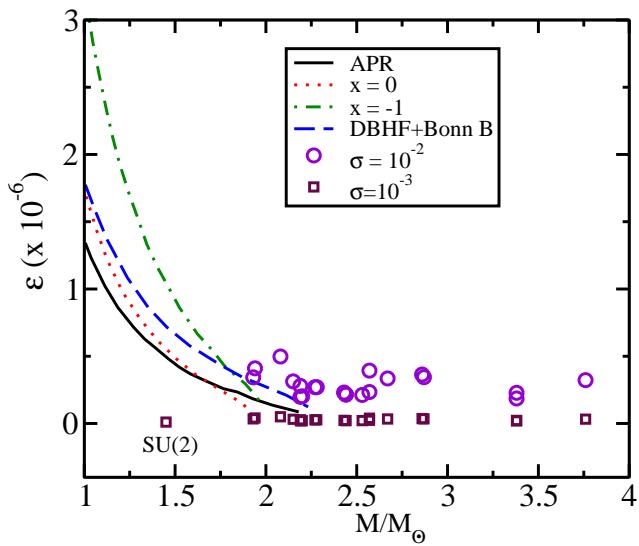}
\caption{Maximum rotational neutron star ellipticity with mass in different 
various parameter sets. }
\label{ell_mass}
\end{figure}

The ellipticity of a neutron star is an important observable, which
gives the structural variation of a star from its spherical shape.
We can calculate it analytically by using equation (\ref{eps}). 
From this equation, 
ellipticity is directly related to the quadrupole moment $\Phi_{22}$ and 
moment of inertia (I) of the NS. 
We have given our calculated results obtained by 
all the 21 force parameters in figure~\ref{ell_mass}. 
We have also compared our results with two theoretical models 
APR (black line) and DBHF + Bonn B (blue dash line) along with the
two experimental results of Ref. \cite{plamen08} for $x=0$ (red dotted line) 
and $-1$ (green dotted dash line). 
Here, we have shown the results of two sets with $\sigma=10^{-2}$ and 
$\sigma=10^{-3}$, which are shown by open circle and square in 
Fig.~\ref{ell_mass}.
As this is rotational star, the maximum mass is larger  
compared to slowly rotating one. If we see the results shown in the 
figure~\ref{ell_mass}, our calculated result still matches with the earlier
work at large NS mass except for SU(2) predictions~\cite{su}. 
Thus our predicted ellipticity of rotating neutron star using various 
parameter sets, where their origin are very different from each 
other are almost similar. 
The variation of the ellipticity (${\epsilon}$) obtained from various star 
mass is very small. This will be helpful for us to constrain the results of 
quadrupole moment, moment of inertia and breaking strain of the neutron star.

\section{Summery and Conclusions}
In this work, we have taken the relativistic and non-relativistic models for 
calculation of gravitational wave strain amplitude, gravitational wave 
frequency, Keplarian frequency, quadrupole moment and ellipticity of 
rotating neutron star. 
We have taken maximum mass and its corresponding radius 
for calculating these observables. Thus, there is an indirect way to 
constraint the maximum mass and radius of the neutron star by these 
observables and vice versa. 
We get almost consistent results in all considered models which show the 
model independent predictions of the observables except for SU(2) 
parameter set. 
We found that gravitational wave strain amplitude 
is a function of breaking strain of neutron star crust and distance between 
the star and the earth. From our calculation, we approximate the
range of the gravitational wave amplitude between $10^{-24}$ to $10^{-22}$ for
rotating neutron star. The moment of inertia of the star comes around $\sim$ 
$10^{45}$ $g cm^2$ and the predicted range of gravitational wave frequency 
is in between $400$ to $1280$ Hz.   
We have calculated the rotating 
frequency of star and 
concluded that, if we increase the rotating frequency then the 
increment in the mass is also changes subsequently. 
The ellipticity of the 
neutron star is consistent in all the considered 21 parameter sets which 
will be helpful to constraint the value of quadrupole and moment of inertia 
of the NS and vice versa.  Our results will be very helpful in the respect 
of the prediction of second and third generation of gravitational wave detector 
family.


\end{document}